\documentclass[]{cJAS2e}

\usepackage{subfigure}
\usepackage{algpseudocode}
\usepackage{algorithm}

\theoremstyle{plain}

\theoremstyle{remark}

\theoremstyle{definition}
\newtheorem{definition}{Definition}

\theoremstyle{example}

\newcommand{\bfx}{\mathbf{x}}                        

\newcommand{\bfX}{\mathbf{X}}                        

\newcommand{\rx}{\mathtt{x}}

\newcommand{\ry}{\mathrm{y}}

\newcommand{\vx}{\mathbf{x}}
\newcommand{\vy}{\mathbf{y}}
\newcommand{\vz}{\mathbf{z}}

\newcommand{\rz}{\mathrm{z}}

\newcommand{\Real}{I \! \! R}
\newcommand{\base}{\mathtt{g}}

\newcommand{\sD}{\mathscr{D}}

\newcommand{\bfR}{\mathbf{R}}
\newcommand{\bfK}{\mathbf{K}}

\usepackage{mathrsfs}
\usepackage{multirow}
\usepackage{multicol} 
\usepackage{booktabs} 

\newcommand{\nmathbf}{\bm}

\def\bfmu{\nmathbf \mu}
\def\bfLambda{\nmathbf \Lambda}
\def\bfepsilon{\nmathbf \epsilon}

\begin{document}


\articletype{}

\title{{\itshape Pattern Discovery in Students' Evaluations of Professors} \break A Statistical Data Mining Approach}

\author{Necla G\"und\"uz$^{\rm a}$$^{\ast}$\thanks{$^\ast$Corresponding author. Email: ngunduz@gazi.edu.tr
\vspace{6pt}} and Ernest Fokou\'e$^{\rm b}$\\\vspace{6pt}  $^{a}${\em{Department of Statistics, Faculty of Science,
Gazi University, Ankara, Turkey}};\\
$^{b}${\em{School of Mathematical Sciences, Rochester Institute of Technology, Rochester, NY 14623, USA}}}

\maketitle

\begin{abstract}
The evaluation of instructors by their students has been practiced at most universities for many decades, and there has always been a great interest in a variety of aspects of the evaluations. Are students matured and knowledgeable enough to provide useful and dependable feedback for the improvement of their instructors' teaching skills/abilities? Does the level of difficulty of the course have a strong relationship with the rating the student give an instructor? In this paper, we attempt to answer questions such as these using some state of the art statistical data mining techniques such support vector machines, classification and regression trees,
boosting, random forest, factor analysis, kMeans clustering. hierarchical clustering. We explore various aspects of the data from both the supervised and unsupervised learning perspective.
The data set analyzed in this paper was collected from a university in Turkey. The application of our techniques to
this data reveals some very interesting patterns in the evaluations, like the strong association between the student's seriousness and dedication (measured by attendance) and the kind of scores they tend to assign to their instructors.


\begin{classcode}\textit{AMS Classification codes}: 62H30; 62H25 \end{classcode}
\end{abstract}

\section{Introduction}
The evaluation of instructors by their students has been practiced at most universities for many decades.
Typically, these evaluations are administered in the form of long surveys answered by students at the end of the semester (quarter).
Questions in the survey are related to such aspects as course organization, level and quality of delivery,
clarity of course objectives, level of difficulty of the course, impact of the course on the student's overall university experience and goals,
relevance of the course, preparedness and competency of the instructor, likeability and fairness of the instructor,
overall satisfaction of the student, and overall rating of the instructor by the student,
just to name a few.

\noindent This study investigates a data set which was anonymously collected in recent years from Gazi University in Ankara (Turkey). It contains a total $5820$ evaluation scores provided by students for three different instructors. There is a total of $28$ course specific questions (see the Appendix for a detailed list of all the questions) presented in Likert-type format, and an additional $3$ attributes, namely {\it student's perceived difficulty level of the course}, {\it attendance}, {\it number of repetitions of the course}.

On the other hand, the overarching goal of students' evaluations of professors is the extraction of knowledge, patterns and information, with the finality of providing their professors with hopefully useful feedback to help them teach better and give students a richer and more effective learning experience. However, there has always been heated debates regarding the effectiveness or even the validity of such evaluations. Many scholars  have wondered over the years if  it is at all possible to improve education quality based on the outcomes of students' evaluations of professors.
For them, they wonder if the answers are informative. Do the answers given by students provide the kind of knowledge and information that can help reshape and improve course quality and professors' teaching abilities?
Typically, most university administrators such as department heads, school directors, college deans,
provosts and chancellors have tended to rely on a single grand average of the questionnaire scores as a measure of the quality of an instructor. \\

\noindent Given the complex and multidimensional nature of the questionnaires administered, it is clearly misleading to summarize such evaluations with a single number. Besides, the averages usually relied upon are not valid, because of the non-numeric nature of the Likert-type of the evaluation responses/scores.\\

\noindent Indeed, since the publication of the seminal \cite{Likert:1932:1} paper, Likert-type scores have been extensively used in a wide variety of fields ranging from Anthropology, Psychology, Education, Sociology, Sports just to name of a few. Unfortunately, with the astronomical number of applications of the Likert measurement system, there have also been innumerable abuses, especially the misuse of Likert-type scores as real-valued scores.\\
\noindent Authors such as \cite{Sisson:1989:1},  \cite{Clason:1994:1}, \cite{Jamieson:2004:1} and \cite{Allen:2007:1} provide pointers to the uses abuses of Likert-type data.
Many authors have indeed cautioned experimenters on the meaninglessness of statements made based on analyses with inappropriate techniques. To quote \cite{Adams:1965:1},  {\it "Nothing is wrong per se in applying any statistical operation to measurements of given scale, but what may be wrong, depending on what is said about the results of these applications, is that the statement about them will not be empirically meaningful or else that it is not scientifically significant"}.
Along the lines of \cite{Adams:1965:1}, many authors have written numerous articles providing guidelines as to which statistical techniques are most appropriate for Likert-type and the so-called
Likert-scale datasets. \cite{Boone:2012:1} of instance provides a clear separation between Likert-type and Likert-scale, and strongly recommends nonparametric
techniques for Likert-type and parametric techniques for Likert-scale.\\

\noindent (Need to correction paragraph)To avoid such pitfalls of meaningless conclusions on our data, we strive to guarantee the validity of our analyses and summaries, by using mostly Likert-type specific (or at least Likert-type compatible) techniques and tools of exploratory data analysis, cluster analysis, dimensionality reduction and pattern recognition.\\

The rest of this paper is organized as follows: in section $2$ we present some general definitions and address important aspects of survey data such item reliability and respondent reliability. We also present empirical answers to most of the above questions using both appropriate exploratory data analysis tools and some straightforward tests of association. In section $3$ we focus on the multivariate aspects of the data and answer most of the students' evaluation of instructors questions by using tools such as factor analytic and cluster analysis which both reveal very meaningful confirmation of some beliefs and perceptions about the rating of professors by their students. Section $4$ uses some of the results from section $3$ to perform predictive analytics on this data. We specifically
apply state of the art pattern recognition techniques such as support vector machines, boosting, random forest and classification trees to predict the satisfaction level of a given student based on their answers to the $28$ questions on the survey. Section $5$ provides our conclusion and discussion, along with pointers to our future work.

\section{Definitions, Data Quality, Exploratory Data Analysis and Basic Tests}
\subsection{Definitions and data quality}
The dataset is represented by an $n \times p$ matrix $\bfX$ whose $i$th row
$\vx_i^\top \equiv (\rx_{i1}, \rx_{i2},\cdots,\rx_{ip})$
denotes the $p$-tuple of characteristics, with each $\rx_{ij} \in \{1,2,3,4,5\}$ representing  the Likert-type level (order) of preference of respondent $i$
on item $j$. Recall that a Likert-type score is obtained by translating/mapping the response levels $\{{\tt Strong \, Disagree}, {\tt Disagree}, {\tt Neutral}, {\tt Agree}, {\tt Strongly \, Agree}\}$ into pseudo-numbers $\{1,2,3,4,5\}$.
\noindent A usually crucial part in the analysis of questionnaire data is the calculation of the Cronbach's alpha coefficient which measures the
reliability/quality of the data. Let $X= (X_1,X_2,\cdots,X_p)^\top$ be a $p$-tuple representing the $p$ items of a questionnaire.
The Cronbach's alpha coefficient is a function of the ratio of the sum of the idiosyncratic item variances over the variance of the sum of the items, and is given by
\begin{eqnarray}
\alpha = \left(\frac{p}{p-1}\right)\left[1-\frac{\sum_{j=1}^p{\mathbb{V}(X_j)}}{\mathbb{V}\left(\sum_{\ell=1}^p{X_\ell}\right)}\right].  \label{eq:cronbach:1}
\end{eqnarray}
For our data, we found $\hat{\alpha}=0.992$, indicating a reliable (ie good quality) survey instrument from Cronbach's point of view. We must emphasize however, that this is just  {\it item reliability}, which is of course of great importance, but herein contrasted with {\it respondent reliability} which we had to assess and address as a result of some patterns discovered in our data.\\


\begin{definition}
Let $\mathcal{D}=\{\vx_1,\vx_2,\cdots,\vx_n\}$ be a dataset with
$\vx_i^\top = (\rx_{i1}, \rx_{i2},\cdots,\rx_{ip})$.
An observation vector $\vx_i$ will be called a {\it zero variation} vector if
$\rx_{ij}= {\tt constant},\,\, j=1,\cdots,p$. Respondents with
{\it zero variation} response vectors will be referred to as
{\tt single minded} respondents/evaluators.
\end{definition}

In our data set of $n=5820$ evaluations, we found a rather high prevalence
of single minded evaluators, specifically,  about half of the evaluations ($2985/5820\approx 51\%$).
In fact, {\it zero variation} responses essentially reduce a $p$ items  survey to a single item survey.


\noindent As can be seen for our earlier calculation, the estimated Cronbach's $\alpha$ coefficient for our data is considerably high. 
The reason may be the high ratio of {\it zero variation} observations. It therefore became interesting to also estimate  the Cronbach's $\alpha$ coefficient for only the {\it non zero variation} observations, which turned out to be $0.9755$. Not surprisingly,  the Cronbach's $\alpha$ value for the {\it zero variation} observations is $1$,  since that corresponds to perfectly reliable  questionnaire. 

{\it Despite the fact that {\it zero variation} responses correspond to a perfectly reliable instrument from Cronbach's alpha perspective,
it is our view that {\it zero variation} responses are an indication that the respondent
did not give deep thought to each of the questions/items of the survey.} One could always argue that such evaluators
came in with a single rating on all the items, and that such responses are just fine, in the sense that {\it they provide a clear and unambiguous
overall assessment of the professor being evaluated.}
However, considering the sometimes drastically different foci
of the questions, it is rather unlikely that a given instructor on a given course would perform exactly the same on all the items. On the other hand, such {\it zero variation} responses convey the impression that the respondent rushed the answering process.
Finally, from a point of view of feedback to the instructor in order to help them improve the course, such answers provide very little if any feedback at all. Authors like  \cite{Marsh:1982:1}, \cite{Stedman:1983:1}, \cite{Marsh:1984:1} and  \cite{Marsh:1997:1} have contributed extended studies and findings related to the effectiveness of students' rating of professors and have also touched extensively on aspects like biases, utility, reliability and validity. In the spirit of most of the points raised by those authors,
we seriously question the effectiveness,  utility, reliability and validity of a students' evaluations data with high incidence/prevalence of zero variation. \cite{Marsh:1982:1}, \cite{Marsh:1983:1}, \cite{Marsh:1984:1}, \cite{Marsh:1992:1} and \cite{Marsh:2007:1} has done
a lot of research work highlighting the importance of adopting a multivariate view of students' evaluations of professors. Clearly, the multivariate aspect of the feedback sought is lost in the prevalence of too many single minded respondent.  For all the above reasons, we deem the {\it zero variation} responses unreliable with respect to the multivariate view of the evaluation.

\begin{definition}
Let $\mathcal{D}=\{\vx_1,\vx_2,\cdots,\vx_n\}$ be a dataset with
$\vx_i^\top = (\rx_{i1}, \rx_{i2},\cdots,\rx_{ip})$. Let the estimated variance of the $i$th respondent be
${\tilde{S}_i^2} = \sum_{j=1}^p{(\rx_{ij}-\bar{\rx}_i)^2/(p-1)}$.
Let $Z_j = \sum_{i=1}^n{\rx_{ij}}$ represent the sum of the scores given by all the $n$
respondents to item $j$. Our respondent reliability is estimated by
\begin{eqnarray}
\hat{\tilde{\alpha}} =
\left(\frac{n}{n-1}\right)\left[1-\frac{\displaystyle \sum_{i=1}^n{\sum_{j=1}^p{\left(\rx_{ij}-\frac{1}{p}\sum_{j=1}^p{\rx_{ij}}\right)^2}}}
{\displaystyle \sum_{j=1}^p{\left(\sum_{i=1}^n{\rx_{ij}}-\frac{1}{p}\sum_{j=1}^p{\sum_{i=1}^n{\rx_{ij}}}\right)^2}}\right].
\label{eq:cronbach:2}
\end{eqnarray}
\end{definition}
We use a straightforward adaptation of the Cronbach's alpha coefficient to measure and capture {\it respondent reliability}.
Given a data matrix $\bfX$, respondent reliability can be computed in practice by simply taking the
Cronbach's alpha coefficient of $\bfX^\top$, the transpose of the data matrix $\bfX$. Let $m$ be the
number of {\it nonzero variation}. If $m \ll p$ and $m/n$ is very small, then respondent reliability will
be very poor. Fortunately, for our data, respondent reliability is estimated at $0.996$, which is very satisfactory.
We think this large value is due to the fact that, despite having more than $50\%$ {\it zero variation} respondents,
we still a large enough sample. Despite this however, we will perform analyses taking into account the dichotomy
between single minded respondents and their counterparts.

\subsection{Exploratory Data Analysis and Basic Tests}
As we said earlier, students' evaluations of instructors are administered with the goal of measuring the effectiveness (quality) of instructors  and hopefully provide them (the instructors) with useful feedback to help them teach better. Clearly, such a goal is complex, and because of its complexity, there have always been heated and often very passionate debates about the validity and the appropriateness of such evaluations \cite{Marsh:1982:1}, \cite{Marsh:1984:1}, \cite{Stedman:1983:1}, \cite{Basow:1995:1}.
As a matter of fact, many professors strongly believe and claim that students, especially undergraduate students, are neither mature enough nor knowledgeable enough nor objective enough to provide useful feedback to their instructors \cite{Feldman:1976:1}, \cite{Abrami:1980:1}, \cite{Peterson:1980:1} and \cite{Marsh:1997:1}.
To a certain degree, such anti-students' evaluations professors do have a valid point because even with the crucial issues of maturity, knowledge and objectivity, there are very important points of concerns with students' evaluations of instructors: (a) a complex multidimensional instrument like a $28$ items questionnaire should never be summarized using a single number (as it is commonly practiced around the world), because such a simplistic summarization definitely fails to capture all the niceties inherent in the complex art of teaching (b) given the Likert type nature of the scores (responses), the often used grand average is at best misleading because averages computed on non-numeric variables are often meaningless  \cite{Adams:1965:1}.
It makes sense that only a multidimensional summary \cite{Marsh:1982:1}, \cite{Marsh:1983:1}, \cite{Marsh:1984:1}, \cite{Marsh:1992:1} and \cite{Marsh:2007:1} or better yet a functional summary (density or mass function) can meaningfully capture the pattern underlying a multidimensional instrument like the students' evaluations of instructors. 

\subsection{Univariate summaries}
It's a very common practice among people dealing with Likert type data to use averages and standard deviations as their measures of central tendency and measures of spread (variation) respectively. Typically, students' evaluations questionnaires have {\bf one} item
aimed at measuring the overall rating of the professor being evaluated.  At universities like Gazi University where the questionnaire does not have such a summarizing item, the grand mean (mean of all the means) is used as the estimate of the overall {\tt rating}, namely
\begin{eqnarray}
{\tt grand mean}(\bfx) = \frac{1}{np}\sum_{j=1}^p{\sum_{i=1}^n{\rx_{ij}}} = {\tt mean}\left(\underset{j=1:p}{\tt mean}(s(\bfx_{j}))\right), 
\label{eq:grandmean:1}
\end{eqnarray}
where $s(\bfx_j) = \{\rx_{ij}: i=1\cdots,n\}$. When an instructor opens the website containing
her/his student evaluation data, there are $28$ averages, one for each questions, and then there is the average of those averages which is the grand mean
representing the overall rating of the instructor. With $\rx_{ij} \in \{1,2,3,4,5\}$, such a
grand average is at best misleading and at worst just plain invalid. In the hierarchy of data types, Likert type scores are no more than ordinal, which prohibits the use of averages. By their very nature, Likert-type observations are inherently definitely not numerical in the usual sense of interval or ratio data.
Considering our motivating example of the students' evaluation of instructors, the use of the grand mean as the overall rating of the instructor misses the subtle and important information revealed by appropriate frequencies (proportions)  and the corresponding bar plots. When the grand mean is used, Instructor 1 scores an average of $3.4$, which of course tells us nothing about the distribution of her scores.
The distribution for this instructor is skewed to the left, with a pronounced/strong mode at $4$ for most of the questions/items. Although we still do not advocate the use of a single number to summarize a complex instrument like a students' evaluations of instructor, we would recommend trusting the mode rather than the mean if a single number were to be used. This led us to defining a grand mode in place of the invalid grand mean as follows:
Let $s(\bfx_j) = \{\rx_{ij}: i=1\cdots,n\}$ and let $\tilde{\rx}_{ij} = {\tt unique}(\rx_{ij})$. If $m_j$ denotes the mode of variable $X_j$, then for $j=1,\cdots,p$, we can readily compute the mode of the $j$th column as
\begin{eqnarray}
m_j = \underset{\tilde{\rx}_{ij}\in s(\bfx_j)}{{\tt arg}\,{\tt max}}\big\{frequency(\tilde{\rx}_{ij})\big\} = {\tt mode}(s(\bfx_{j})).
\label{eq:grandmode:1}
\end{eqnarray}
\noindent The set $M = \{m_1, m_2,\cdots, m_p\}$ containing the  modes for the $p$ columns. Let $\tilde{m}_j = {\tt unique}(m_{j})$.
We can find the grand mode as
\begin{eqnarray}
{\tt grand mode}(\bfx) = \underset{\tilde{m}_{j} \in M}{{\tt arg}\,{\tt max}}\big\{frequency(\tilde{m}_{j})\big\} = {\tt mode}\left(\underset{j=1:p}{\tt mode}(s(\bfx_{j}))\right). 
\label{eq:grandmode:2}
\end{eqnarray}
\noindent The grand mode for instructor $1$ is found to be $4$, which, in light of the distribution of her scores, is a more accurate summarization of her effectiveness and teaching quality. It might be tempting, given the ordinal nature of Likert-type data, to use the grand median
\begin{eqnarray}
{\tt grand median}(\bfx) = {\tt median}\left(\underset{j=1:p}{\tt median}(s(\bfx_{j}))\right), 
\label{eq:grandmedian:1}
\end{eqnarray}
in place of the grand mean. From our experience, such a summarization is not as accurate as the grand mode, partly due to the floor and ceiling effect, see \cite{Clason:1994:1}.
If instead of considering only instructor $1$ we use the entirety of the data with all the $n=5820$ evaluations, the distribution of all the $28$ course specific questions, it is then noted that most questions attain their mode at $3$, and we find the grand mode to be $3$.
Thanks to the distributional features of the scores of the instructors in this dataset, namely the skewness to the left, we able to comment in a more complete manner on the effectiveness (or at least the students' perception thereof). With the highest frequencies being between 3 and 5, it is fair to say that the instructors evaluated here are {\bf not} negatively perceived by their students.

\subsubsection{Examining the Effect of Response Variation}
Despite the ability to provide a more meaningful single summarization of the whole evaluation through the grand mode along with distributional qualifications, we still need to answer relational questions like the association between
student maturity and their rating, student seriousness/dedication/objectivity and their rating. We now propose to focus on {\it zero variation} responses, as we believe that the reflect the reliability of the respondent. In a sense, we claim that a student who gives a {\it zero variation} response is providing a less objective and less mature answer to the survey. We then try to find out if there is an association between {\it zero variation} and the answers to the questions.
First and foremost, it is interesting to assess the association between response variation and instructors. See Table \ref{tab:zerovar:1}.
\begin{table}[!htbp]
\centering
\begin{tabular}{lccc}
\toprule
    & {\tt Zero Variation} & {\tt Nonzero Variation} & {\tt Total}\\   \hline
{\tt Instructor 1}  & 0.0789 & 0.0543 & 0.1332\\
{\tt Instructor 2}  & 0.1309 & 0.1172 & 0.2481\\
{\tt Instructor 3}  & 0.3031 & 0.3156 & 0.6187\\ \hline
{\tt Total} &  0.5129 & 0.4871 & {\bf 1} \\
\bottomrule
\end{tabular}
\caption{Distribution of response variation among instructors}
\label{tab:zerovar:1}
\end{table}

The chi-squared test of association between {\it Instructor} and {\it Response Variation} is significant, namely with $\chi_{\tt obs}^2 = 28.45$, ${\tt df} = 2$, {\it p-value} $ = 0.000$.  This means that there are some differences among instructors with respect to zero variation responses.\\

The lack of richness of the {\it zero variation} responses in this study is less concerning because most of such responses
are either neutral or positive. In a sense, those who were single minded about their rating of the courses, were so mostly not because of dissatisfaction. Given the fact that the zero variation responses came from satisfied students (see higher percentage of Neutral, Agree, and Strongly Agree in Table \eqref{tab:zerovar:2} depicting the percentages of responses within the zero variation group),  their feedback was really not needed with respect to the multidimensional aspect of the feedback sought. For that reason, we can proceed with the remaining aspects of the analysis of this data, secured that both the item reliability (measured by Cronbach's alpha) and the respondent reliability are satisfactory.
\begin{table}[!h]
\centering
\begin{tabular}{lccccc}
\toprule
    & {\tt Strongly Disagree} & {\tt Disagree} & {\tt Neutral} & {\tt Agree} & {\tt Strongly Agree}\\   \hline
{\tt Instr 1}  & 0.1678 & 0.0545 & 0.2288 & 0.2767 & 0.2723 \\
{\tt Instr 2}  & 0.1378 & 0.0407 & 0.3018 & 0.3084 & 0.2113 \\
{\tt Instr 3}  & 0.2086 & 0.0726 & 0.3294 & 0.2183 & 0.1712 \\
\bottomrule
\end{tabular}
\caption{Proportion of each response category for answers with zero variation}
\label{tab:zerovar:2}
\end{table}

\subsubsection{Examining Various Important Associations}
We now examine a variety of association between different important variables. Taking the view that {\it attendance}
is a measure of dedication/seriousness, and therefore a decent and plausible indicator of the ability/authority of the student to correctly assess their instructor, we will now test the association of various variables with {\it attendance}. In other words, if a student is not dedicated ie not serious (as measured by attendance), their assessment should probably not be taken seriously. See \cite{Marsh:1981:3} for a detailed account on the influence of the student's interest on their rating of their instructor. We also consider the variable {\it difficulty}, a self reported variable provided to allow the student to indicate their perception of the level of difficulty of the course.
This variable is particularly important because some instructors strongly believe that students tend to give a negative feedback when they perceive the course to be difficult. Extensive studies on the impact of the difficulty level of the course being evaluated have been carried out by authors such as \cite{OConnell:1993:1} and \cite{Clayson:2009:1}.\\

{\it Association between Attendance Level and Response Variation for Instructor 3.}
\begin{table}[!htbp]
\label{tab:zerovar:attendance:3}
\centering
\begin{tabular}{lccccc}
\toprule
    & {\tt Poor} & {\tt Minimal} & {\tt Good} & {\tt Good} & {\tt Excellent}\\   \hline
  {\tt nonzero} & 0.14912524 & 0.08747570 & 0.07220217 & 0.12663149 & 0.07470147 \\
  {\tt zero}    & 0.22299361 & 0.07497917 & 0.05942794 & 0.07275757 & 0.05970564 \\
\bottomrule
\end{tabular}
\caption{Cross tabulation of Attendance level vs Response Variation for Instructor 3.}
\end{table}

\noindent As can be seen on Table \eqref{tab:zerovar:attendance:3}, there is empirical evidence of a substantial difference between zero variation and nonzero variance respondent in the group with poor attendance. The corresponding chi-squared test of association between {\it Attendance level} and {\it Response Variation} is significant, specifically with $\chi_{\tt obs}^2 = 117.7398$, $\nu={\tt df} = 4$, {\it p-value} $ < 2.2\times 10^{-16}$.\\

{\it Association between Difficulty Level and Response Variation for Instructor 3.}

\begin{table}[!htbp]
\centering
\begin{tabular}{lccccc}
\toprule
    & {\tt Too Easy} & {\tt Easy} & {\tt Normal} & {\tt Difficult} & {\tt Too Difficult}\\   \hline
  {\tt nonzero} & 0.12385448 & 0.04498750 & 0.14940294 & 0.12968620 & 0.06220494 \\
  {\tt zero}    & 0.19550125 & 0.03471258 & 0.10274924 & 0.09080811 & 0.06609275 \\
\bottomrule
\end{tabular}
\caption{Cross tabulation of Difficulty level vs Response Variation given by students for Instructor 3.}
\label{tab:zerovar:difficulty:3}
\end{table}

\noindent In the group of those who deemed the course to be too easy, there appears to be some evidence of more zero variation respondents. More formally, the corresponding chi-squared test of association between {\it Difficulty level} and {\it Response variation} is significant, specifically with
$\chi_{\tt obs}^2 = 117.7398$, $\nu={\tt df} = 4$, {\it p-value} $ < 2.2\times 10^{-16}$.  \\

{\it Various Tests of Association using the whole dataset} \\

It appears that for all the evaluations provided for Instructor $3$, both {\it Attendance Level}
{\it Difficulty Level} are strongly associated with {\it Response Variation}. The question then arises as to whether that association holds when all the $5820$ evaluations are considered.

\begin{table}[!htbp]
\label{tab:zerovar:attendance:all}
\centering
\begin{tabular}{lccccc}
\toprule
    & {\tt Poor} & {\tt Minimal} & {\tt Good} & {\tt Good} & {\tt Excellent}\\   \hline
  {\tt nonzero} & 0.1273 & 0.0880 & 0.0718 & 0.1218 & 0.0782 \\
  {\tt zero}    & 0.1995 & 0.0887 & 0.0643 & 0.0933 & 0.0672 \\
\bottomrule
\end{tabular}
\caption{Cross tabulation of Attendance level vs Response Variation for all the Instructors.}
\end{table}

\noindent The corresponding chi-squared test of association between {\it Attendance level} and {\it Response variation} is significant, namely with
$\chi_{\tt obs}^2 = 118.3$, $\nu={\tt df} = 4$, {\it p-value} $ < 2.2\times 10^{-16}$

\begin{table}[!htbp]
\label{tab:zerovar:difficulty:all}
\centering
\begin{tabular}{lccccc}
\toprule
    & {\tt Too Easy} & {\tt Easy} & {\tt Normal} & {\tt Difficult} & {\tt Too Difficult}\\   \hline
  {\tt nonzero} & 0.1065 & 0.0527 & 0.1601 & 0.1158 & 0.0519 \\
  {\tt zero}    & 0.1718 & 0.0416 & 0.1447 & 0.0947 & 0.0601 \\
\bottomrule
\end{tabular}
\caption{Cross tabulation of Difficulty level vs Response Variation for all the Instructors.}
\end{table}

\noindent The corresponding chi-squared test of association between {\it Difficulty level} and {\it Response variation} is significant, namely with $\chi_{\tt obs}^2 = 113.5$, $\nu={\tt df} = 4$, {\it p-value} $ < 2.2\times 10^{-16}$. \\

Students with poor attendance give an overwhelmingly large number of {\it zero variation} answers whereas students with reasonable to excellent attendance level tend to give {\it nonzero variation} answers. This somewhat confirms or at least supports the strongly held belief that only those answers provided by dedicated/serious students should be taken into account.  On the other hand,
students who perceive a course as too easy and therefore boring or at least uninteresting also tend to give an overwhelming proportion of {\it zero variation} answers. Interestingly, students who think the course has a normal difficulty level tend to take time to provide varied answers to different questionnaire items.

\begin{figure}[!htbp]
\centering
\epsfig{figure=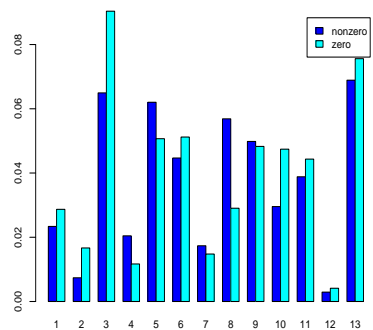, height=5.5cm, width=5.5cm}
\caption{Cross-tabulation of response variation versus course indicator for all the instructors.}
\label{fig:barplot:zero:course:all}
\end{figure}

 Since we discovered interesting patterns between response variation and both attendance and difficulty level, it is interesting to examine if there might be a similar type of strong association between the courses and the response variation.
 Indeed many other authors have researched on the relationship between the nature of the courses taught and the rating of the instructors. See \cite{Marsh:1981:3} and \cite{Marsh:1983:1} for more information.
 In our dataset, there was a total of $13$  courses included in the $5820$ evaluations considered. Figure \eqref{fig:barplot:zero:course:all}
 depicts the barplot of the relationship between response variation and the courses. It can be seen that except for five courses, zero variation respondents are the majority.
The chi-squared test of association between {\it Course} and {\it Response variation} was found to be significant, with
$\chi_{\tt obs}^2 = 150.7$, $\nu={\tt df} = 12$, {\it p-value} $ < 2.2\times 10^{-16}$.

Finally, we look at the overall relationship between attendance and the perceived difficulty level of the course. Table \eqref{tab:attendance:difficulty:all} shows an overwhelming support in favor of a strong relationship, with dominance of the strength between too easy and poor. 

\begin{table}[!htbp]
\centering
\begin{tabular}{lccccc}
\toprule
    & {\tt Too Easy} & {\tt Easy} & {\tt Normal} & {\tt Difficult} & {\tt Too Difficult}\\   \hline
  {\tt Poor} & 0.2263 & 0.0170 & 0.0380 & 0.0249 & 0.0206 \\
  {\tt Minimum}    & 0.0137 & 0.0311 & 0.0653 & 0.0431 & 0.0234\\
  {\tt Reasonable}    & 0.0093 & 0.0131 & 0.0591 & 0.0393 & 0.0153\\
  {\tt Good}    & 0.0158 & 0.0187 & 0.0859 & 0.0679 & 0.0268\\
  {\tt Excellent}    & 0.0132 & 0.0144 & 0.0565 & 0.0352 & 0.0259\\
\bottomrule
\end{tabular}
\caption{Cross tabulation of Difficulty level vs Attendance Level for all the Instructors.}
\label{tab:attendance:difficulty:all}
\end{table}

\noindent The corresponding chi-squared test of association between {\it Difficulty level} and {\it Attendance Level} is significant, with
$\chi_{\tt obs}^2 = 2528.06$, $\nu={\tt df} = 16$, {\it p-value} $ < 2.2\times 10^{-16}$. The most obvious feature of this association is the astronomically high proportion of poor attendance in courses deemed too easy. No surprise here, just plain common sense. Sadly however, there is no category in which excellent attendance dominates. \\

Although all the $28$ items in the questionnaire were carefully selected by the designers of the students' evaluation, one could make a strong case that some questions are better indicators of overall assessment than others. 
One such question is {\it Q10:\,My initial expectations about the course were met at the end of the period or year.}
This question is in fact often used as the measure of the overall assessment of the course and the instructor at most American universities.
The University of Central Florida students' evaluation questionnaire given in Appendix has a question phrased in a very similar way.
Given its summarizing nature, we will use this question as a response/dependent variable in our supervise learning section.  

%

\section{Pattern Recognition and Association Analysis}
In this section, we turn our attention to the multivariate aspects of our data set. We perform Factor Analysis to extract
meaningful latent structure and cluster analysis to identify potential groups in the way students rate their professors. 
\cite{OConnell:1993:1}, \cite{Hsu:2007:1}, \cite{Wang:2009:1}, \cite{Badur:2011:1}, \cite{Ola:2013:1} and \cite{Syed:2014:1} 
are some of the authors who have used data mining and machine learning techniques on student evaluation data.
As we said clearly in sections 1 and 2, Likert-type scores are inherently non-numeric, and applying techniques designed for 
numeric data on Likert-type will yield answers that are potentially meaningless or at best very difficult to interpret. When it comes to correlation
analysis for instance, the default choice is the Pearson correlation measure. With Likert type data however,
one wonders if Pearson correlation should ever be used. Based on recommendations by \cite{Adams:1965:1}, \cite{Boone:2012:1},
and \cite{Clason:1994:1}, the correct type of correlation for Likert-type data should be Kendall-tau-B correlation
or the Spearman correlation, as these are designed for (ordinal) ranked data.
There have been many recent interesting contributions to the multivariate analysis of Likert-type:
in her doctoral thesis, \cite{Javaras:2004:1} provides a wide variety of univariate and multivariate tools for analyzing Likert-type data.
\cite{Narli:2010:1} proposes the use of rough sets in the analysis of Likert-scale data.
We start off by checking how different the Pearson correlation matrix would be from the Kendall-tau B correlation matrix on our data.
Recall, that given two random variables $X_i$ and $X_j$ for which observed (realized) values $\rx_{1i}, \rx_{2i}, \cdots , \rx_{ni}$ and
$\rx_{1j}, \rx_{2j}, \cdots , \rx_{nj}$ have been respectively gathered, the so-called Pearson sample
correlation matrix is given by
$$
r_{ij} = {\tt correlation}(\rx_i,\rx_j) = r(\rx_i,\rx_j) =
\frac{1}{n-1}\sum_{\ell=1}^n{\left(\frac{\rx_{\ell i}-\bar{\rx}_i}{s_{\rx_i}}\right)\left(\frac{\rx_{\ell j}-\bar{\rx}_j}{s_{\rx_j}}\right)}.
$$
For a random $p$-tuple $X = (X_1,X_2,\cdots,X_p)^\top$ and the corresponding data matrix $\bfX = (\rx_{ij}), i=1,\cdots,n, \, j=1,\cdots,p$,
the Pearson sample correlation matrix is given by
$$
\bfR = \left[\begin{array}{cccc}
1 & r_{12} & \cdots & r_{1p}\\
r_{21} & 1 & \cdots & r_{2p}\\
\vdots & \vdots & \ddots & \vdots\\
r_{p1} & r_{p2} & \cdots & 1\\
\end{array}
\right].
$$
As we have been stressing all along, the Likert-type nature of our data makes the matrix $\bfR$ meaningless, in the sense that the averages on which it is based may not have an interpretable meaning. If we consider two Likert type (ordered categorical) variables $X$ and $Y$ once again, their Kendall $\tau${-}B correlation coefficient $\tau_B(X,Y)$ is given by
$$
\tau_B(X,Y) = \frac{n_c(X,Y) - n_d(X,Y)}{\sqrt{(n_0-n_1)(n_0-n_2)}},
$$
where $n_0=n(n-1)/2$, $n_1=\sum_i{t_i(t_i-1)/2}$, $n_2=\sum_j{u_j(u_j-1)/2}$,
$t_i$={\it number of tied values in the i-th group of ties for the first quantity},
$u_j$={\it number of tied values in the j-th group of ties for the second quantity},
$n_c$ = {\it number of concordant pairs}, $n_d$={\it number of discordant pairs}.
For a $p$-tuple  $X=(X_1,\cdots,X_p)$ of $p$ Likert type variables, the Kendall Tau-B correlation matrix is $\bfK$ where
$$
\bfK = \left[\begin{array}{cccc}
\tau_{B}(X_1,X_1) & \tau_{B}(X_1,X_2) & \cdots & \tau_{B}(X_1,X_p)\\
\tau_{B}(X_2,X_1) & \tau_{B}(X_2,X_2) & \cdots & \tau_{B}(X_2,X_p)\\
\vdots & \vdots & \ddots & \vdots\\
\tau_{B}(X_p,X_1) & \tau_{B}(X_p,X_2) & \cdots & \tau_{B}(X_p,X_p)\\
\end{array}
\right].
$$

The empirical calculations based on our data reveal that the Pearson and the Kendall $\tau$-B correlation matrices are so similar, in pattern and magnitude as to be almost indistinguishable (in fact, it turns out that $\widehat{\bfR} \approx \widehat{\bfK} + 0.05 I_p$)\footnote{This strong similarity is somewhat surprising because while Pearson is appropriate for numeric data types, Kendall $\tau$-B is suitable for ordinal and rank like Likert data in our study.}. For that reason, in our subsequent correlation-based analyses like Principal Component Analysis or Factor Analysis, we can use the Pearson in place of the Kendall $\tau$-B correlation, since the latter is computationally very expensive.

\subsection{Are there distinguishable groups among students?}
A natural question that arises in the presence of data like the one we have, is whether the observations can be clustered.
In other words, is there such a thing as different groups of students as far as their patterns of feedback to instructors are concerned?
Can the patterns of students' evaluations of their instructors be grouped into distinct and clearly describable categories?
Now, one of the most celebrated approaches to cluster analysis is the ubiquitous kMeans clustering algorithm\footnote{
In a previous analysis of the dataset used in this paper,  \cite{gunduz:Likert:2013:1} explored
various of cluster analysis, including hierarchical clustering on both the raw data and transformed versions of the data
for which the Jaccard distance was used. More details of that analysis can be found through the reference.}.
Obviously, as the name suggests, it is based on the computation of averages that represent the centers of potential
underlying groups. With Likert-type data, it has been stressed all along that averages are potentially meaningless because of 
the inherently categorical non numeric nature of such data. With the Pearson and Kendall-tau-B correlation matrices computed earlier
showing strong similarities in pattern and magnitude, one might conjecture that 
it might not be wrong to use average-driven techniques on our data.
The kMeans clustering algorithm in this case would proceed by partitioning the data into $k$ clusters to form the 
optimal partitioning 
$\mathcal{P}^* = C_1^* \cup C_2^* \cup \cdots \cup C_k^*$ that minimizes the within-cluster
sum of squares (WCSS). In other words, if $\mathcal{P}^*$ denotes the best partitioning (clustering) of the data, we must have
$$
\mathcal{P}^* = {\tt arg}\underset{\mathcal{P}}{\tt min}\left\{\sum_{j=1}^k{\sum_{i=1}^n{I(\bfx_i \in C_j)\|\bfx_i-\bfmu_j\|^2}}\right\}
$$
where $\bfmu_j$ is the mean vector (center) for cluster $C_j$. From our kMeans clustering calculations in R,
the percentage of variation explained seems to clearly suggest that one should retain three distinct clusters.
Indeed, two clusters would capture a very low percentage of the variation in the data,
while four clusters do not substantially improve the amount of variation captured by three clusters. We therefore
retained three clusters and carefully examined both the percentage of observations in each one of them and
the values of the centers. As Table \ref{tab:cluster:results:1} shows, one could venture to say that
almost $60\%$ of the students have a neutral opinion of the courses they took, and this seems to apply
to almost all the $28$ questions of the survey. The cluster analytic result also suggests that $17\%$
expressed maximum satisfaction with the courses they took. Finally a third group of the students
seems to be the group of very dissatisfied students, with our data showing roughly $23\%$ of such students.
These numbers apply to all the $5820$ evaluations analyzed. It certainly would be more beneficial,
in the interest of instructor's improvement, to extract such clustering for each course in order to
help the instructor identify areas of improvement.
\begin{table}[!htbp]
\centering
\begin{tabular}{lccc}
\toprule
    & {\tt Cluster 1} & {\tt Cluster 2} & {\tt Cluster 3}\\   \hline
{\tt Average of Center}  & 4.80 & 1.52 & 3.37\\
{\tt Number of Observations}  & 1010 & 1364 & 3446\\
{\tt Percentage of observations}  & 17.35\% & 23.44\% & 59.21\% \\ \hline
{\tt Suggested class label}  & {\tt Satisfied} &  {\tt Dissatisfied} &  {\tt Neutral} \\
\bottomrule
\end{tabular}
\caption{Clusters extracted using kMeans clustering.}
\label{tab:cluster:results:1}
\end{table}

The patterns discovered through kMeans clustering and revealed in Table \eqref{tab:cluster:results:1}
are interesting in their own right, but as we'll show later, we used the labels generated here to 
extend our analysis to supervised learning. As we mentioned earlier, the dataset  \cite{GunduzFokoue:2013:1} used here was gathered 
at Gazi University where there is no dedicated response variable in the questionnaire.  
Given this absence of response, we later use  $Y \in \{ \mathtt{Dissatisfied}, \mathtt{Neutral}, \mathtt{Satisfied}\}$ 
as our response variable in both classification trees and random forest.

\subsection{Are there meaning concepts underlying the items of the evaluation?}
It goes without saying that $28$ questions for a single respondent can be quite overwhelming. Besides, it's indeed very likely
that many of the questions end measuring the same aspect of the perception of the student. Recall for instance that the correlation
matrices calculated earlier revealed extremely large correlation values. We should therefore expect the $28$ dimensional questionnaire
given to students to boil down to a much lower number of latent concepts.  From a factor analytic perspective, this means that the
student evaluation vector $\vx^\top=(\rx_1,\cdots,\rx_{28})$ does have a representation of the form
$$
\vx = \bfLambda \vz  + \bfepsilon
$$
where $\bfLambda \in \Real^{28 \times q}$ and $\vz^\top=(\vz_1,\cdots,\vz_{q})$ for some $q \ll 28$. Factor Analysis typically assumes that
the factor scores vector $Z$ has a multivariate Gaussian (normal) distribution. Such an assumption is bound to be violated here because of the
non-normality of the vector $X$. Many authors have performed factor analysis on Likert-type data despite this non-normality.
\cite{Muthen:1992:1} and \cite{Lubke:2004:1} provide a detail account of the pitfalls resulting from the misuses of factor analysis on Likert-type data.
It turns out that part of the problem with the use of factor analysis on Likert-type data stems from the fact that some analysts use the Pearson covariance matrix as their main ingredient. To somehow avoid the pitfalls and hope for meaningful factor analytic results, we use the Kendall $\tau$-B correlation matrix as the basis of our factor analysis. Based on Table \ref{tab:factanal:1} our factor analytic seem to reveal the following facts: Questions $13$ to $28$ have estimated factor loadings that are all higher on factor $1$ than they are on Factor $2$. These $16$ questions
are all related to how the student rate the competence of the instructor teaching the course.
We therefore name the first factor score $Z_1$ the "{\bf instructor rating score}".
Questions $1$ to $12$ have estimated factor loadings that are all higher on factor $2$ than they are on Factor $1$. These $12$ questions
are all related to how satisfied the student was about the course. We therefore name the second factor score $Z_2$ the "{\bf student satisfaction score}".

\begin{table}[!ht]
  \centering
  \subtable[Revelation of Factor 2\label{tab:factanal:1a}]{
  \centering
  \begin{tabular}{lrr}
        \toprule
            & {\tt Factor 1} & {\tt Factor 2} \\   \hline
        Q1  & 0.376   & \bf{0.781} \\
        Q2  & 0.495   & \bf{0.767} \\
        Q3  & 0.567   & \bf{0.689} \\
        Q4  & 0.475   & \bf{0.770} \\
        Q5  & 0.505   & \bf{0.793} \\
        Q6  & 0.497   & \bf{0.776} \\
        Q7  & 0.465   & \bf{0.819} \\
        Q8  & 0.456   & \bf{0.815} \\
        Q9  & 0.545   & \bf{0.699} \\
        Q10 & 0.524   & \bf{0.791} \\
        Q11 & 0.564   & \bf{0.680} \\
        Q12 & 0.486   & \bf{0.751} \\
            &         &       \\
            &         &       \\
            &         &       \\
            &         &       \\
        \bottomrule
        \end{tabular}
  }
  \subtable[Revelation of Factor 1\label{tab:factanal:1b}]{
    \centering
    \begin{tabular}{lrr}
        \toprule
            & {\tt Factor 1} & {\tt Factor 2} \\   \hline
        Q13 & \bf{0.753}   & 0.558 \\
        Q14 & \bf{0.794}   & 0.517 \\
        Q15 & \bf{0.791}   & 0.514 \\
        Q16 & \bf{0.705}   & 0.611 \\
        Q17 & \bf{0.827}   & 0.391 \\
        Q18 & \bf{0.762}   & 0.541 \\
        Q19 & \bf{0.790}   & 0.517 \\
        Q20 & \bf{0.825}   & 0.475 \\
        Q21 & \bf{0.844}   & 0.447 \\
        Q22 & \bf{0.846}   & 0.446 \\
        Q23 & \bf{0.756}   & 0.564 \\
        Q24 & \bf{0.713}   & 0.593 \\
        Q25 & \bf{0.826}   & 0.463 \\
        Q26 & \bf{0.749}   & 0.543 \\
        Q27 & \bf{0.695}   & 0.567 \\
        Q28 & \bf{0.811}   & 0.452 \\
   \bottomrule
   \end{tabular}
   }
  \caption{Two-factor model from the $p=28$ questions on the Gazi University students' evaluation data. There was a total of $n=5820$ evaluations submitted by the students and used to estimate these factor loadings. It can be seen that the two factors discovered are quite straightforward.}
  \label{tab:factanal:1}
\end{table}

The two factors described above captured $85\%$ of the variation, and any attempt to generate/derive more factors resulted in
very little gain along with the loss of interpretability inherent in these two factors. From a practical perspective,
it seems to make sense that a student's answers would be summarized into their overall satisfaction along with some rating of the instructor
who led the whole experience on the course. Clearly one could hypothesize more factors, but these two tend to intuitively
capture what one would expect. To help better grasp the usefulness of the factor analytic patterns that we discovered
in this data, we deem it appropriate to match the initial scores given by some students with the corresponding factor scores. 
For instance, 
\begin{itemize}
  \item {\it Evaluator satisfied with both course and professor: } the factor scores for student $25$ were found to be $\vz_{25}=(0.84, 1.40)$, revealing that this student was satisfied with the course overall ($\rz_{25,2}=1.40>0$) and also satisfied with the instructor's organization and running of the course ($\rz_{25,1}=0.84>0$). Indeed, we also found the input from this student to be
$\vx_{25}=(5,5,5,5,5,5,5,5,5,5,5,5, 5,  5,  5,  5,  5,  5,  5,  5,  5,   5,   5,   5,   5,   5,   5,   5)$, which confirms, based on their empathically high scores, that they are doubly satisfied.
  \item {\it Evaluator strongly dissatisfied with both course and professor: } A quick look at student number $96$'s initial scores shows a candidate doubly dissatisfied with both the course and the professor, namely with a zero variation input vector given by 
$\vx_{96}=(1,1,1,1,1,1,1,1,1,1,1,1, 1,  1,  1,  1,  1,  1,  1,  1,  1,   1,   1,   1,   1,   1,   1,   1)$. 
The factor scores for this person confirm it with two negative numbers, namely $\vz_{96}=(-1.60, -1.06)$.
  \item {\it Evaluator strongly dissatisfied with course but neutral with professor: } A perfect example in this group is epitomized by student $541$
  with an input vector given by $\vx_{541}=(1, 1, 1, 1, 1, 1, 1, 1, 1, 1, 1, 1,3,   3,   3,   3,   3,   3,   3,   3,   3,   3,   3,   3,   3,   3,   3,   3)$, and corresponding factor scores  $\vz_{541}=(1.60, -2.95)$. The initial scores from this student seem to indicate a strong dissatisfaction
  with the course and rather neutral view of all the aspects related to the professor. One could almost imagine this candidate saying: "{\it I hate this course and got nothing out of it, and I really don't have anything for or against the professor who taught it}".
  \item {\it Evaluator strongly satisfied with course but strongly dissatisfied with professor: } Looking at this vector
$\vx_{474}=(5,5,5,5,5,5,5,5,5,5,5,5, 1,  1,  1,  1,  1,  1,  1,  1,  1,   1,   1,   1,   1,   1,   1,   1)$, it's clear that student
$574$ was strongly dissatisfied with every single aspect of the professor, but still came away from the course with a strong satisfaction.
The factor model summarizing this student's evaluation succinct in  $\vz_{474}=(-5.44, 5.15)$.
  \item {\it Evaluator neutral about  both course and professor: }  With an input vector given by 
$\vx_{123}=(3,3,3,3,3,3,3,3,3,3,3,3, 3, 3,3,3,3,3,3,3,3,3,3,3,3,3,3,3)$,  student $123$ is a typical case of an evaluator who was lukewarm about every aspect of both the course and professor. The factor model captures this quite well with $\vz_{123}=(-0.38,  0.16)$.
\end{itemize}

The above specific examples were provided to further show evidence that the 2-factor model
extracted the latent concepts underlying the Gazi University students' evaluation instrument rather well.
As we saw from all the above calculations and findings, one can readily estimate the satisfaction of a given student
and their rating of their professor using two numbers, the factor scores, and in this case, 
the numbers are clear and unambiguous.

\section{Supervised Learning Techniques}
Until this point, all our analyses on this data have been entirely unsupervised. While we have discovered 
many interesting patterns in the data, we are now turning to supervised learning with the hope of discovering even
more interesting aspects of how students rate their professors. Throughout this section, we'll concentrate
on classification using trees and random forests. For all our random forest estimations we'll use $500$ trees
in the ensemble. Since the data came without a specifically dedicated response variable, we'll use {\tt Q10}
as one of our response variable for reasons mentioned earlier. We will use another response variable herein denoted
by {\tt Opinion}, whose domain contains the labels
 {\tt Dissatisfied}, {\tt Neutral}, {\tt Satisfied} generated earlier from kMeans clustering.
 
\subsection{Classification Tree Learning}
\begin{figure}
\begin{center}
\subfigure[Classification tree with $Y=Opinion$. \label{fig:tree:Opinion:Q10:class:regr:1a}]{
\resizebox*{6.5cm}{!}{\includegraphics{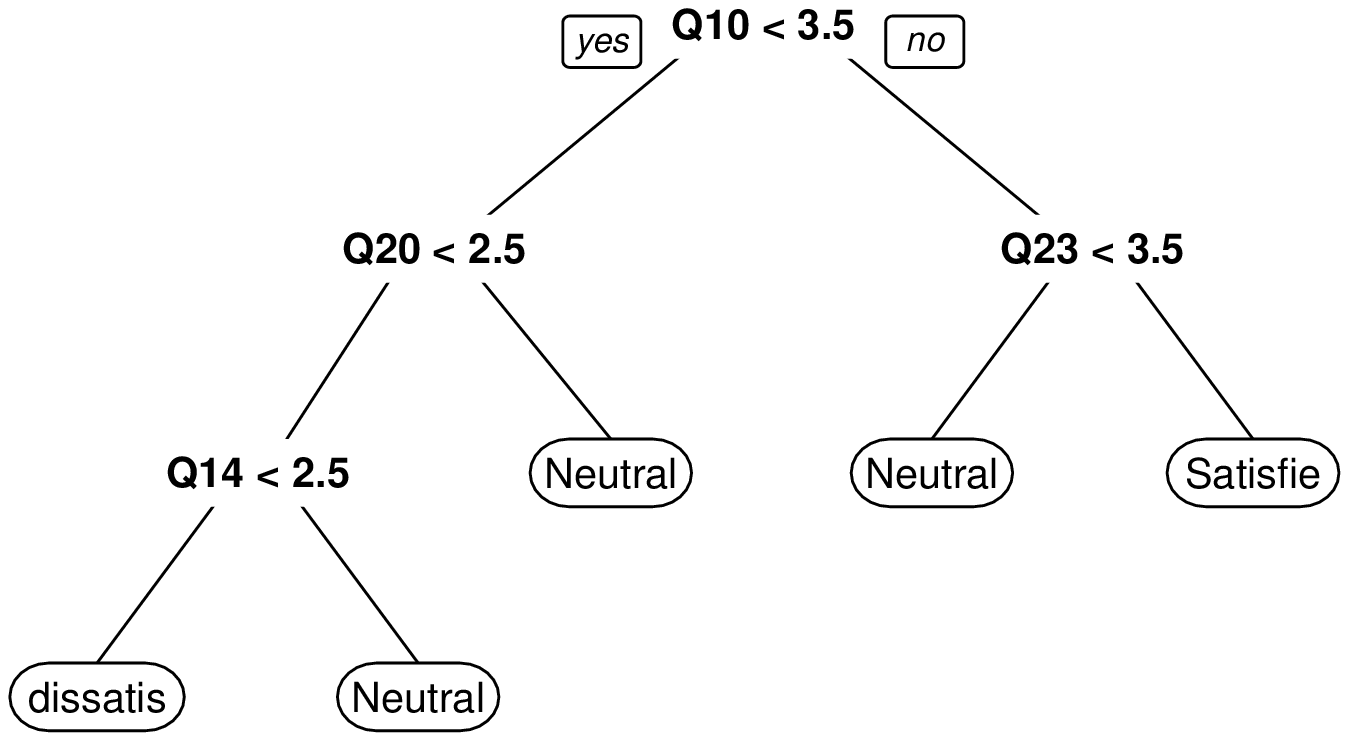}}}\hspace{6pt}
\subfigure[Classification tree with $Y=Q10$.\label{fig:tree:Opinion:Q10:class:regr:1b}]{
\resizebox*{6.5cm}{!}{\includegraphics{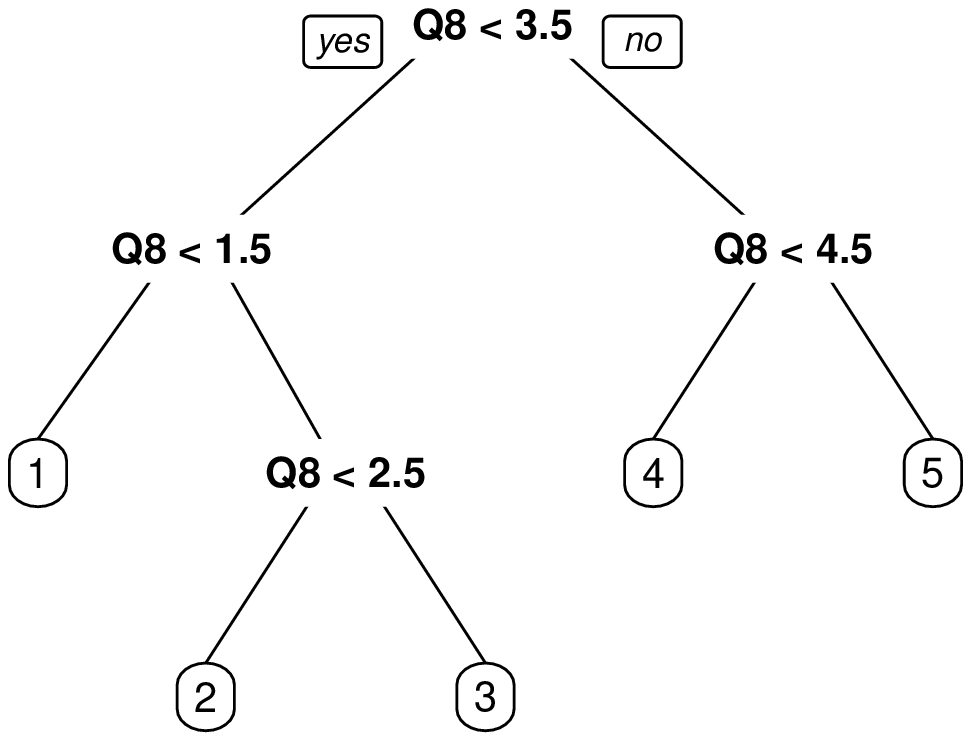}}}
\caption{Classification Trees with two different Response Scenarios.}
\label{fig:tree:Opinion:Q10:class:regr:1}
\end{center}
\end{figure}

 The great appeal of trees was triggered by our interest in finding out if there were some questions that drove the classification and that could therefore be considered somewhat key questions in the evaluation.
Classification trees are usually highly preferred by analysts who desire an interpretable learning machine. Understanding trees is indeed straightforward as they are intuitively appealing piecewise functions operating on a partitioning of the input space. Given $\mathscr{D} = \Big\{(\bfx_1, Y_1),  \cdots, (\bfx_n, Y_n)\Big\}$, with $\bfx_i \in \mathscr{X}$, $Y_i \in \{1,\cdots,G\}$.
Let $T = \displaystyle \cup_{\ell=1}^q{R_\ell}$ denote the tree represented by the partitioning of $\mathscr{X}$ into $q$ regions $R_1, R_2, \cdots, R_q$. Given a new point $\bfx^*$, its predicted response is
$$
\hat{Y}_{\tt Tree}^* = \hat{f}_{\tt Tree}(\bfx^*) =  \sum_{\ell=1}^q\left\{I(\vx^* \in R_\ell)\left\{\underset{j \in \{1,\cdots,G\}}{\tt argmax} \left\{\frac{1}{|R_\ell|}\sum_{\bfx_i \in R_{\ell}}{I(Y_i = j)}\right\}\right\}\right\}.
$$
The classification tree software implementation used here is taken from the R package {\sf rpart}. Using 
the labels  {\tt Dissatisfied}, {\tt Neutral}, {\tt Satisfied} as our response levels, we get the tree depicted in Figure  
\eqref{fig:tree:Opinion:Q10:class:regr:1a}, which clearly reveals variable Q10 as the root, somewhat lending support to our 
earlier speculation around the possibility of using Q10 as the response variable because of its apparent summarizing nature.
On Figure \eqref{fig:tree:Opinion:Q10:class:regr:1b}, we depict the classification tree generated using Q10 as the response.
Bearing in mind the instability of classification trees (high estimation variance), we dedicate the last subsection of this paper
to an extension of tree learning, namely the ubiquitous ensemble learning method known as Random Forest \cite{Breiman:2001:1},
which we apply to our data in the next section. One extra motivation for resorting to ensemble learning via Random Forest lies 
in our desire to estimate how well the rating of a professor can be predicted, but also to estimate the importance of each of the variables
used in the prediction.

\subsection{Ensemble Learning with Random Forests}
Let's consider once again the multi-class classification task as defined much earlier
with labels $\ry$ coming from $\mathscr{Y}=\{1,2,\cdots,G\}$ and predictor variables
 $\vx=(\rx_1,\rx_2,\cdots,\rx_p)^\top$  coming from a $p$-dimensional space $\mathscr{X}$.
Along the same lines of \cite{Gunduz:Fokoue:2015:1}, let $\widehat{\tt g}^{\tt (b)}(\cdot)$ be the $b$th bootstrap replication of the estimated base classifier $\widehat{\tt g}(\cdot)$,
such that $(\widehat{\ry})^{(b)} = \widehat{\tt g}^{\tt (b)}(\vx^*)$ is the $b$th bootstrap  estimated class of $\vx^*$. 
The estimated response by Random Forest is obtained using the majority vote rule, which means that the most frequent label throughout the $B$ bootstrap replications of random subspace learning. The following algorithmic description taken from \cite{Gunduz:Fokoue:2015:1}
captures the essential structure of the Random Forest\cite{Breiman:2001:1} learning method.

\begin{algorithm}
\caption{Random Subspace Learning for Model Aggregation}\label{algo:rf:1}
\begin{algorithmic}[1]
\Procedure{RandomForest}{$B$}\Comment{The Random Forest Algorithm for $B$ trees}
\State {\tt Choose a base learner $\widehat{\base}(\cdot)$} \Comment{e.g.:  Trees}
\State {\tt Choose an estimation method} \Comment{e.g.:  Recursive Partitioning}
\For{$b=1$ to $B$}
\State {\tt Draw with replacement} from $\sD$ a bootstrap sample $\sD^{(b)} = \{\vz_1^{(b)},\cdots,\vz_n^{(b)}\}$
\State {\tt Draw without replacement} from $\{1,2,\cdots,p\}$ a subset $\{j_1^{(b)},\cdots,j_d^{(b)}\}$ 
\State {\tt Drop unselected variables} from $\sD^{(b)}$ so that $\sD_{\tt sub}^{(b)}$ is $d$ dimensional
\State {\tt Build the $b$th base learner} $\widehat{\base}^{(b)}(\cdot)$ based on $\sD_{\tt sub}^{(b)}$
\EndFor \label{algo:rf:1}
\State Use the ensemble $\Big\{\widehat{\base}^{(b)}(\cdot),\, b=1,\cdots, B\Big\}$, the predicted label of $\vx^*$ is
\begin{eqnarray*}
\widehat{f}^{\tt (RF)}(\vx^*) = {\tt arg} \,\underset{\ry \in \mathscr{Y}}{\tt max}\left\{{\tt freq}_{\widehat{\tt C}^{\tt (B)}(\vx^*)}(\ry)\right\}
= {\tt arg} \,\underset{\ry \in \mathscr{Y}}{\tt max}\left\{\sum_{b=1}^{B}{\left({\bf 1}_{\{\ry=\widehat{\tt g}^{\tt (b)}(\vx^*)\}}\right)}\right\}.
\end{eqnarray*}
\EndProcedure
\end{algorithmic}
\end{algorithm}
Table \eqref{tab:confmat:rf:opinion:1} depicts the confusion matrix of the $500^{th}$ random tree of the
forest built using  Opinion as the response. The last column is actually the training error and should not be mistaken for the average test error that measures the generalization ability of random forest. The R Package {\sf randomForest} automatically gives the out of bag (OOB) error for each random tree.  In the spirit of \cite{Gunduz:Fokoue:2015:1}, we shall use the average OOB error  $\mathtt{AVOOB}(\cdot)$, as our measure of predictive performance, namely
\begin{eqnarray}
    \label{eq:avoob:1}
    \mathtt{AVOOB}(\widehat{f}^{\tt (RF)}) =\frac{1}{B} \sum_{b=1}^{B} \left\{ \frac{1}{m} \sum_{i=1}^{m} \ell(\vy_{i}^{(b)}, \widehat{\base}^{(\tt b)}(\vx_i^{(b)}))\right\},
\end{eqnarray}
where the observations $\{(\vx_i^{(b)}, \ry_i^{(b)}), i=1,\cdots,m\}$ are the $m \approx \lceil e n\rceil$ observations not selected by
the bootstrap sampling with replacement process.

\begin{table}[!htbp] \centering
\begin{tabular}{@{\extracolsep{5pt}} lrrrr}
\toprule
 & {\tt Dissatisfied} & {\tt  Neutral} & {\tt Satisfied} & {\tt class.error} \\
\hline \\[-1.8ex]
{\tt  Dissatisfied} & $1,219$ & $20$ & $0$ & $0.016$ \\
{\tt  Neutral} & $25$ & $2,306$ & $27$ & $0.022$ \\
{\tt  Satisfied} & $0$ & $31$ & $2,192$ & $0.014$ \\
\botrule
\end{tabular}
\caption{Confusion Matrix corresponding to random forest classification with {\tt Opinion} as the response
variable.}
  \label{tab:confmat:rf:opinion:1}
\end{table}
The average OOB error obtained from the above forest comes out as $0.02115943$, meaning that the Random Forest
classifier is $98\%$ accurate. We also built a 500-tree random forest with Q10 as the response. Table
\eqref{tab:rf:confusion:Q10:1} shows the corresponding confusion matrix, and our average out of bag error
for this random forest is found to be $0.1389903$, which interesting corresponds to an accurary of $86\%$,
the percentage of variation captured by the two-factor factor analytic model discovered earlier,
and also the percentage of variation that warranted the selection of the $3$ clusters solution in our kMeans
clustering analysis.

\begin{table}[!htbp] \centering
\begin{tabular}{@{\extracolsep{5pt}} rrrrrrr}
\toprule
 & 1 & 2 & 3 & 4 & 5 & {\tt Class Error} \\
\hline \\[-1.8ex]
1 & $862$ & $63$ & $15$ & $1$ & $1$ & $0.085$ \\
2 & $44$ & $594$ & $135$ & $14$ & $0$ & $0.245$ \\
3 & $10$ & $109$ & $1,521$ & $123$ & $8$ & $0.141$ \\
4 & $1$ & $18$ & $138$ & $1,248$ & $36$ & $0.134$ \\
5 & $1$ & $4$ & $14$ & $63$ & $797$ & $0.093$ \\
\botrule
\end{tabular}
\caption{Confusion Matrix corresponding to random forest classification with {\tt Q10} as the response
variable.}
\label{tab:rf:confusion:Q10:1}
\end{table}

\begin{figure}
\begin{center}
\subfigure[Random Forest with Opinion as response. \label{fig:rf:varImp:class:1a}]{
\resizebox*{7.0cm}{!}{\includegraphics{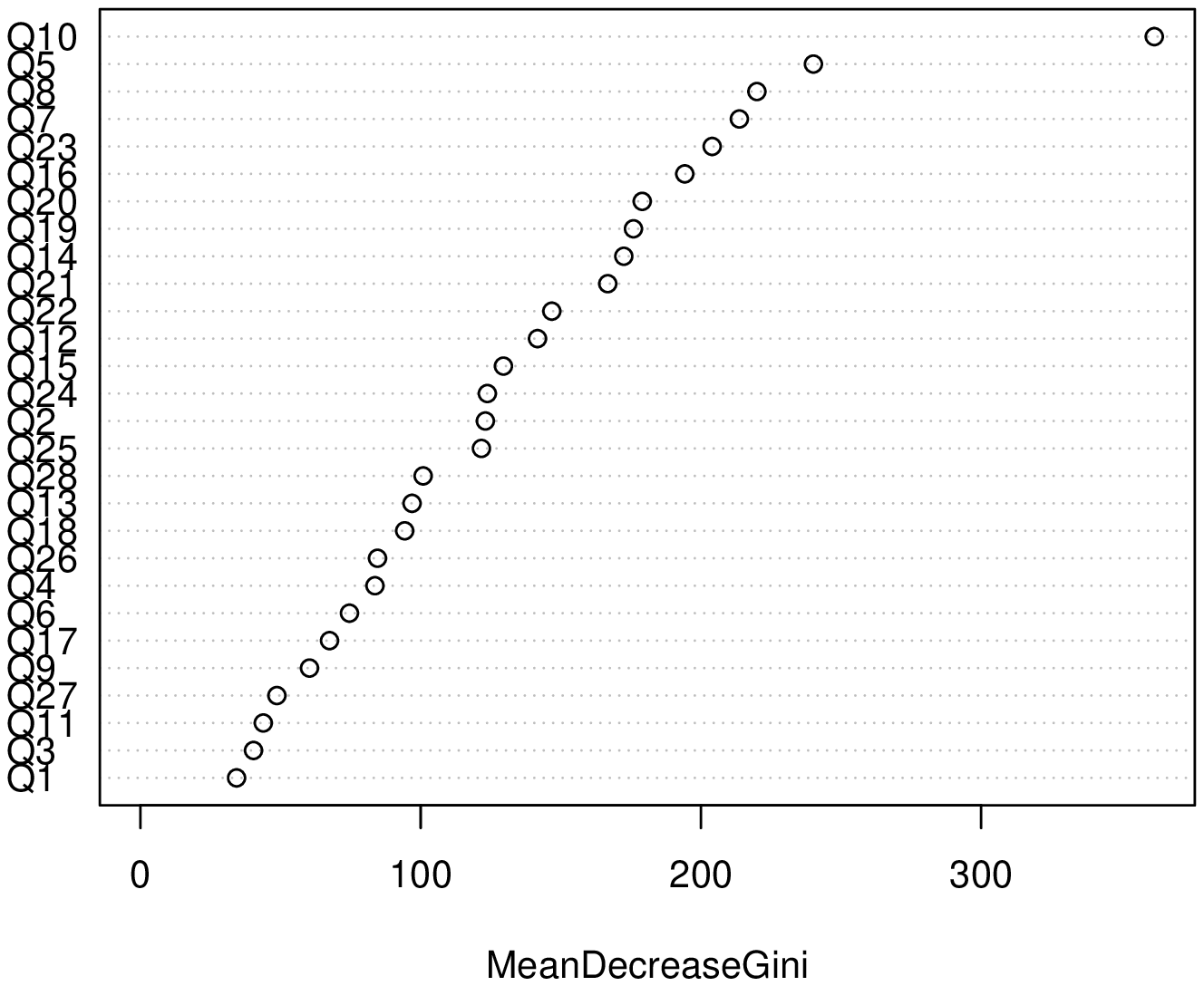}}}\hspace{6pt}
\subfigure[Random Forest with Q10 as response. \label{fig:rf:varImp:class:1b}]{
\resizebox*{7.0cm}{!}{\includegraphics{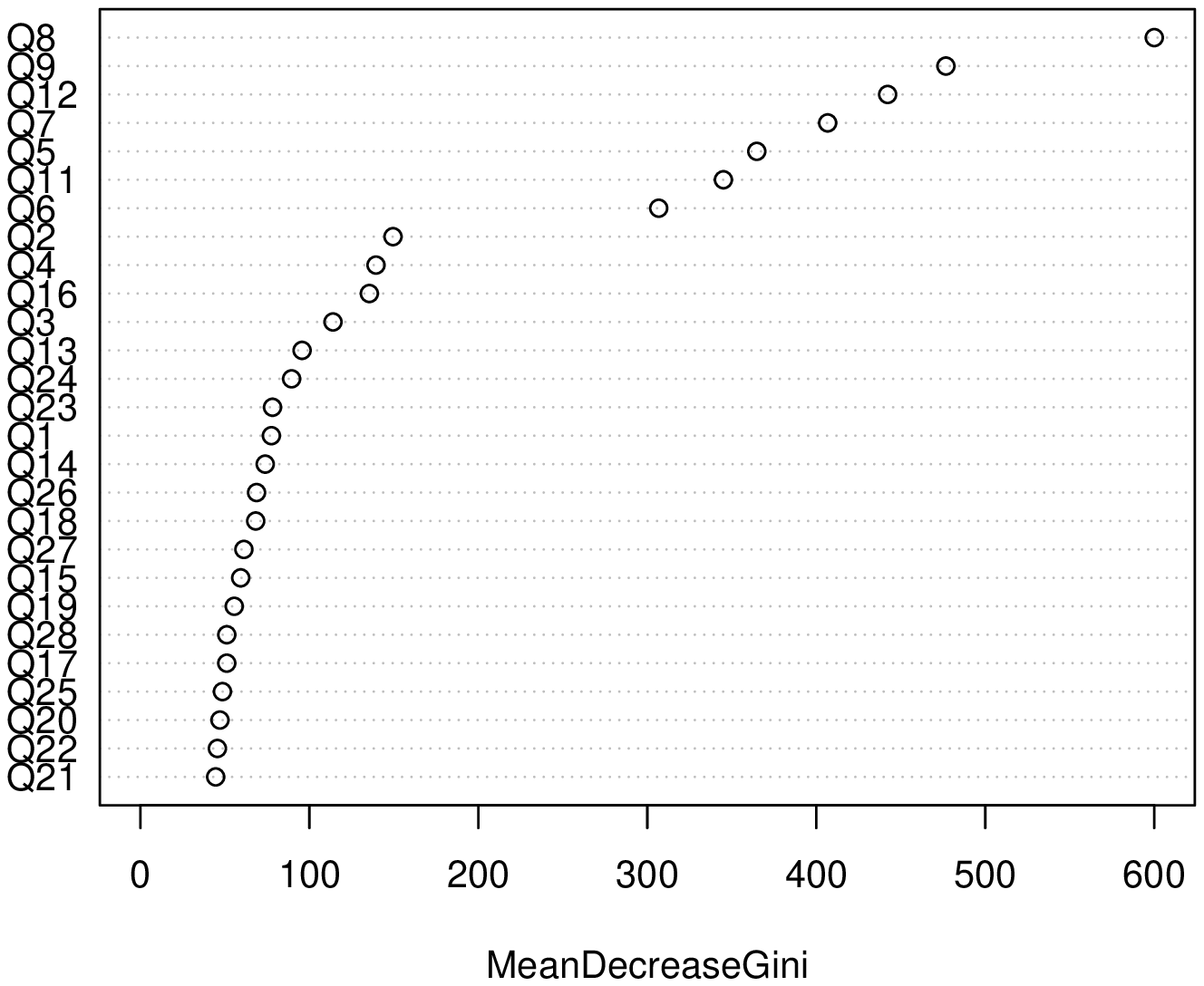}}}
\caption{Variable Importance plots yielded by Random Forest using different scenarios of response variable.}
\label{fig:rf:varImp:class:1}
\end{center}
\end{figure}

One of the greatest appeals of Random Forest lies in its ability to supplement excellent
predictions with estimates of variable importance. For our dataset, we generated two different  variable importance
plots \eqref{fig:rf:varImp:class:1}, one using Q10 as the response and the other using Opinion as the response. 
Figure \eqref{fig:rf:varImp:class:1a} shows the overwhelming dominance of Q10, again confirming our initial 
conjecture/speculation. Interestingly, out of the $5$ most important variables in this case, only one, namely the 5th, is
 related to the professor. On the other hand, Figure \eqref{fig:rf:varImp:class:1b} reveals that to directly predict Q10 well based
on the remaining $27$ variables, the most important variable is Q8, dominating the rest substantially. Shockingly,
for the prediction of Q10, virtually no professor-related variable (Q13-Q28) appear to be important. It gives the impression
that student satisfaction has nothing to do with the professor. Interestingly also, Q10 is almost perfectly predicted by Q8 which has to
do wıth the grades student had on the course. {\it Could it be then that students don't care about who the professor is, as long as they end up 
with a good grade on the course?} 

\section{Conclusion and Discussion}
We have provided a comprehensive statistical analysis of a relatively large dataset containing
students' evaluations of various courses at a university in Turkey.
Factor Analytic results appear to reveal a very plausible two factor model suggesting that
students' evaluations inherently reveal the overall satisfaction of the student at the end of the course
along with impact the instructor had on their overall satisfaction. With the instructor's factor
coming out as the most dominant one, it is fair to say that the instructor does play a central
role in the over experience of the student.
Anyone analyzing students' evaluations should be careful to consider the number of {\it zero variation} responses
and examining their association with the pattern of answers provided by the students. We strongly believe that these
{\it zero variation} responses somewhat determine the quality of the survey and reliability of the answers provided.
We have shown evidence to support the fact dedicated students (attendance) will tend to reveal a more satisfactorily learning
experience than those students who do not take their course seriously.
We combined unsupervised and supervised learning techniques and were able, not only to find meaningful
and interpretable groups in the data, but also identify the items in the questionnaire that appeared to be driving the
students' assessment of their learning experience.The dominance of {\tt Question 10} on the three class recognition tree confirmed our intuition in the sense
that it is the question that seems to measure the overall satisfaction of a student on a course, and it is
re-assuring to have the tree model reveal it.
From a questionnaire design perspective, it is our view that $28$ questions is a bit too much for the students,
and this usually large survey might be the reason why some students ended up giving {\it zero variation} responses.
We would also like to suggest the use of two questions that have been found to be very revealing of the experience of student, namely
(1) {\it What is your overall rating of this instructor?} (2) {\it Would you recommend this course to any other student?}.
Although these two questions are inherently correlated part of our future work on this data will consist of adapting traditional 
classification trees to Likert-type data. 
This essentially boils down to using Likert-type specific loss functions for splitting the nodes of the tree.
We specifically plan on deriving adaptations of  the Jaccard distance as loss function or using the cross entropy measure on the
tendencies of respondents.
We are also planning to include the final grades of the respondents. The motivation for this is the fact that many instructors around the world
have repeatedly argued that students who know (based on quiz scores, homework assignment scores, and midterm exam scores) that they will be receiving a good grade on the course tend to rate their professors very highly. We plan on finding out if there is evidence to support such a belief.

\subsection{Acknowledgements}
Ernest Fokou\'e wishes to express his heartfelt gratitude and infinite thanks to Our Lady of Perpetual Help for Her
ever-present support and guidance, especially for the uninterrupted flow of inspiration received through Her
most powerful intercession.

\bibliographystyle{chicago}
\bibliography{gf-ordinal-1}

\begin{thebibliography}{}

\bibitem[\protect\citeauthoryear{Abrami, Dickens, Perry, and Leventhal}{Abrami
  et~al.}{1980}]{Abrami:1980:1}
Abrami, P.~C., W.~J. Dickens, R.~P. Perry, and L.~Leventhal (1980).
\newblock Do teacher standards for assigning grades affect student evaluations
  of teaching.
\newblock {\em Journal of Educational Psychology\/}~{\em 72\/}(1), 107--118.

\bibitem[\protect\citeauthoryear{Adams, Fagot, and Robinson}{Adams
  et~al.}{1965}]{Adams:1965:1}
Adams, E., R.~F. Fagot, and R.~E. Robinson (1965).
\newblock A theory of appropriate statistics.
\newblock {\em Psychometrica\/}~{\em 30\/}(2), 99--127.

\bibitem[\protect\citeauthoryear{Allen and Seaman}{Allen and
  Seaman}{2007}]{Allen:2007:1}
Allen, I. and C.~A. Seaman (2007).
\newblock Likert scales and data analyses.
\newblock {\em Quality Progress.\/}~{\em 47}, 417--442.

\bibitem[\protect\citeauthoryear{Badur and Mardikyan}{Badur and
  Mardikyan}{2011}]{Badur:2011:1}
Badur, B. and S.~Mardikyan (2011).
\newblock Analyzing teaching performance of instructors using data mining
  techniques.
\newblock {\em Informatics in Education\/}~{\em 10\/}(2), 245–257.

\bibitem[\protect\citeauthoryear{Basow}{Basow}{1995}]{Basow:1995:1}
Basow, S.~A. (1995, December).
\newblock Student evaluations of college professor: When gender matters.
\newblock {\em Journal of Educational Psychology\/}~{\em 87\/}(4), 656--665.

\bibitem[\protect\citeauthoryear{Boone and Boone}{Boone and
  Boone}{2012}]{Boone:2012:1}
Boone, H. and A.~Boone (2012).
\newblock Analyzing likert data.
\newblock {\em Journal of Extenson\/}~{\em 50\/}(2).

\bibitem[\protect\citeauthoryear{Breiman}{Breiman}{2001}]{Breiman:2001:1}
Breiman, L. (2001).
\newblock Random forests.
\newblock {\em Machine Learning\/}~{\em 45\/}(1), 5--32.

\bibitem[\protect\citeauthoryear{Clason and Dormody}{Clason and
  Dormody}{1994}]{Clason:1994:1}
Clason, D. and T.~Dormody (1994).
\newblock Analyzing data measured by individual likert type items.
\newblock {\em Journal of Agricultural Education.\/}~{\em 33\/}(4), 31--35.

\bibitem[\protect\citeauthoryear{Clayson}{Clayson}{2009}]{Clayson:2009:1}
Clayson, D.~E. (2009).
\newblock Student evaluations of teaching: are they related to what students
  learn? a meta-analysis and review of the literature.
\newblock {\em Journal of Marketing Education\/}~{\em 31\/}(1), 16–30.

\bibitem[\protect\citeauthoryear{Feldman}{Feldman}{1976}]{Feldman:1976:1}
Feldman, K.~A. (1976).
\newblock Grades and college students' evaluations of their courses and
  teachers.
\newblock {\em Research in Higher Education\/}~{\em 4}, 69--111.

\bibitem[\protect\citeauthoryear{G\"und\"uz and E.}{G\"und\"uz and
  E.}{2013}]{gunduz:Likert:2013:1}
G\"und\"uz, N. and F.~E. (2013, November).
\newblock Data mining and machine learning techniques for extracting patterns
  in students' evaluations of instructors.
\newblock Technical Report 1746, Rochester Institute of Technology, Rochester,
  New York, USA.

\bibitem[\protect\citeauthoryear{Gunduz and Fokoue}{Gunduz and
  Fokoue}{2013}]{GunduzFokoue:2013:1}
Gunduz, N. and E.~Fokoue (2013).
\newblock {UCI} machine learning repository.

\bibitem[\protect\citeauthoryear{Gunduz and Fokoue}{Gunduz and
  Fokoue}{2015}]{Gunduz:Fokoue:2015:1}
Gunduz, N. and E.~Fokoue (2015, January).
\newblock {Robust Classification of High Dimension Low Sample Size Data}.
\newblock {\em arXiv.org\/}~{\em stat.AP}.

\bibitem[\protect\citeauthoryear{Hsu, Chang, and Hung}{Hsu
  et~al.}{2007}]{Hsu:2007:1}
Hsu, C., B.~Chang, and H.~Hung (2007, December).
\newblock Applying svm to build supplier evaluation model - comparing likert
  scale and fuzzy scale.
\newblock In {\em 2007 IEEE International Conference on Industrial Engineering
  and Engineering Management}, Singapore, pp.\  6--10. IEEE.

\bibitem[\protect\citeauthoryear{Jamieson}{Jamieson}{2004}]{Jamieson:2004:1}
Jamieson, S. (2004).
\newblock Likert scales: How to (ab)use them.
\newblock {\em Medical Education.\/}~{\em 38\/}(38), 1212--1218.

\bibitem[\protect\citeauthoryear{Javaras}{Javaras}{2004}]{Javaras:2004:1}
Javaras, K.~N. (2004, Hilary Term).
\newblock {\em Statistical Analysis of Likert Data on Attitudes}.
\newblock Ph.d. thesis, Balliol College, University of Oxford.

\bibitem[\protect\citeauthoryear{Likert}{Likert}{1932}]{Likert:1932:1}
Likert, R. (1932, June).
\newblock A technique for the measurement of attitudes.
\newblock {\em Archives of Psychology\/}~{\em 22\/}(140), 5--55.

\bibitem[\protect\citeauthoryear{Lubke and Muthen}{Lubke and
  Muthen}{2004}]{Lubke:2004:1}
Lubke, G. and B.~Muthen (2004).
\newblock Factor-analyzing likert-scale data under the assumption of
  multivariate normality complicates a meaningful comparison of observed groups
  or latent classes.
\newblock {\em Structural Equation Modeling\/}~{\em 11\/}(514-534).

\bibitem[\protect\citeauthoryear{Marsh and Cooper}{Marsh and
  Cooper}{1981}]{Marsh:1981:3}
Marsh, H. and T.~Cooper (1981).
\newblock Prior subject interest, students' evaluation, and instructional
  effectiveness.
\newblock {\em Multivariate Behavioral Research\/}~{\em 16}, 82--104.

\bibitem[\protect\citeauthoryear{Marsh}{Marsh}{1982}]{Marsh:1982:1}
Marsh, H.~W. (1982).
\newblock Validity of students' evaluations of college teaching a
  multitrait-multimethod analysis.
\newblock {\em Journal of Educational Psychology\/}~{\em 74\/}(2), 264--279.

\bibitem[\protect\citeauthoryear{Marsh}{Marsh}{1983}]{Marsh:1983:1}
Marsh, H.~W. (1983, February).
\newblock Multidimensional ratings of teaching effectiveness by students from
  different academic settings and their relation to student/course/instructor
  characteristics.
\newblock {\em Journal of Educational Psychology\/}~{\em 75\/}(1), 150--166.

\bibitem[\protect\citeauthoryear{Marsh}{Marsh}{1984}]{Marsh:1984:1}
Marsh, H.~W. (1984, Oct).
\newblock Students' evaluations of university teaching: Dimensionality,
  reliability, validity, potential biaises, and utility.
\newblock {\em Journal of Educational Psychology\/}~{\em 76\/}(5), 707--754.

\bibitem[\protect\citeauthoryear{Marsh}{Marsh}{2007}]{Marsh:2007:1}
Marsh, H.~W. (2007).
\newblock {\em The Scholarship of Teaching and Learning in Higher Education: An
  Evidence-Based Perspective}, Chapter Student's evaluations of university
  teaching: A multidimensional perspective, pp.\  319--384.
\newblock New York: Springer.

\bibitem[\protect\citeauthoryear{Marsh and Dunkin}{Marsh and
  Dunkin}{1992}]{Marsh:1992:1}
Marsh, H.~W. and M.~J. Dunkin (1992).
\newblock {\em Higher education: Handbook of theory and research}, Volume~8,
  Chapter Student's evaluations of university teaching: A multidimensional
  perspective, pp.\  143--233.
\newblock New York: Agathon Press.

\bibitem[\protect\citeauthoryear{Marsh and Roche}{Marsh and
  Roche}{1997}]{Marsh:1997:1}
Marsh, H.~W. and L.~A. Roche (1997, November).
\newblock Making students' evaluations of teaching effectiveness effective: The
  critical issues of validity, bias, and utility.
\newblock {\em American Psychologist\/}~{\em 52\/}(11), 1187--1197.

\bibitem[\protect\citeauthoryear{Muthen and Kaplan}{Muthen and
  Kaplan}{1992}]{Muthen:1992:1}
Muthen, B. and D.~Kaplan (1992).
\newblock A comparison of some methodologies for the factor analysis of
  non-normal likert variables: A note on the size of the model.
\newblock {\em British Journal of Mathematical and Statistical
  Psychology\/}~{\em 45\/}(19-30).

\bibitem[\protect\citeauthoryear{Narli}{Narli}{2010}]{Narli:2010:1}
Narli, S. (2010, March).
\newblock An alternative evaluation method for likert type attitude scales:
  Rough set data analysis.
\newblock {\em Scientific Research and Essays\/}~{\em 5\/}(6), 519--528.

\bibitem[\protect\citeauthoryear{OConnell and Dickinson}{OConnell and
  Dickinson}{1993}]{OConnell:1993:1}
OConnell, D.~Q. and D.~J. Dickinson (1993).
\newblock Student ratings of instruction as a function of testing conditions
  and perceptions of amount learned.
\newblock {\em Journal of Research and Development in Education\/}~{\em
  27\/}(1), 18--23.

\bibitem[\protect\citeauthoryear{Ola and Pallaniappan}{Ola and
  Pallaniappan}{2013}]{Ola:2013:1}
Ola, A.~F. and Pallaniappan (2013).
\newblock A data mining model for evaluation of instructors' performance in
  higher institutions of learning using machine learning algorithms.
\newblock {\em International Journal of Conceptions on Computing and
  Information Technology\/}~{\em 1\/}(2), 17--22.

\bibitem[\protect\citeauthoryear{Peterson and Cooper}{Peterson and
  Cooper}{1980}]{Peterson:1980:1}
Peterson, C. and S.~Cooper (1980).
\newblock Teacher evaluations by graded and ungraded students.
\newblock {\em Journal of Educational Psychology\/}~{\em 72\/}(5), 682--685.

\bibitem[\protect\citeauthoryear{Sisson and Stoker}{Sisson and
  Stoker}{1989}]{Sisson:1989:1}
Sisson, D. and H.~Stoker (1989).
\newblock Analyzing and interpreting likert-type survey data.
\newblock {\em The Delta Pi Epsilon Journal.\/}~{\em 3\/}(2), 81--85.

\bibitem[\protect\citeauthoryear{Stedman}{Stedman}{1983}]{Stedman:1983:1}
Stedman, C. (1983).
\newblock The reliability of teaching effectiveness rating scale for assessing
  faculty performance.
\newblock {\em Tennessee Education\/}~{\em 12\/}(3), 25--32.

\bibitem[\protect\citeauthoryear{Syed, Jiang, and Golab}{Syed
  et~al.}{2014}]{Syed:2014:1}
Syed, S.~J., Y.~H. Jiang, and L.~Golab (2014).
\newblock Data mining of undergraduate course evaluations.
\newblock In {\em Proceedings of the 7th International Conference on
  Educational Data Mining}, pp.\  347--348.

\bibitem[\protect\citeauthoryear{Wang, Dziuban, Cook, and Moskal}{Wang
  et~al.}{2009}]{Wang:2009:1}
Wang, M.~C., C.~D. Dziuban, I.~J. Cook, and P.~D. Moskal (2009).
\newblock {\em Quality Research in Literacy and Science Education}, Chapter Dr.
  Fox Rocks: Using Data-mining Techniques to Examine Student Ratings of
  Instruction, pp.\  383--398.
\newblock Springer.

\end{thebibliography}

\section{Appendices}\label{appendices}
\subsection{Appendix}\label{appendix-gazi}
Student questionnaire from Gazi University:
{\small
\begin{itemize}
  \item Q1:  The semester course content, teaching method and evaluation system were provided at the start.
  \item Q2:  The course aims and objectives were clearly stated at the beginning of the period.
  \item Q3:  The course was worth the amount of credit assigned to it.
  \item Q4:  The course was taught according to the syllabus announced on the first day of class.
  \item Q5:	 The class discussions, homework assignments, applications and studies were satisfactory.
  \item Q6:  The textbook and other courses resources were sufficient and up to date.				
  \item Q7:  The course allowed field work, applications, laboratory, discussion and other studies.
  \item Q8:  The quizzes, assignments, projects and exams contributed to helping the learning.	
  \item Q9:  I greatly enjoyed the class and was eager to actively participate during the lectures.
  \item Q10: My initial expectations about the course were met at the end of the period or year.
  \item Q11: The course was relevant and beneficial to my professional development.
  \item Q12: The course helped me look at life and the world with a new perspective.
  \item Q13: The Instructor's knowledge was relevant and up to date.
  \item Q14: The Instructor came prepared for classes.
  \item Q15: The Instructor taught in accordance with the announced lesson plan.
  \item Q16: The Instructor was committed to the course and was understandable.
  \item Q17: The Instructor arrived on time for classes.
  \item Q18: The Instructor has a smooth and easy to follow delivery/speech.
  \item Q19: The Instructor made effective use of class hours.
  \item Q20: The Instructor explained the course and was eager to be helpful to students.
  \item Q21: The Instructor demonstrated a positive approach to students.
  \item Q22: The Instructor was open and respectful of the views of students about the course.
  \item Q23: The Instructor encouraged participation in the course.
  \item Q24: The Instructor gave relevant homework assignments/projects, and helped/guided students.
  \item Q25: The Instructor responded to questions about the course inside and outside of the course.
  \item Q26: The Instructor's evaluation system (midterm and final questions, projects, assignments, etc.) effectively measured the course objectives.
  \item Q27: The Instructor provided solutions to exams and discussed them with students.
  \item Q28: The Instructor treated all students in a right and objective manner.
\end{itemize}
}

\subsection{Appendix}\label{appendix-florida}
Student perception of instruction items for the University of Central Florida:\\
Source Questions Administration
\begin{enumerate}
  \item Feedback concerning your performance in this course was:
  \item The instructor’s interest in your learning was:
  \item Use of class time was:
  \item The instructor’s overall organization of the course was:
  \item Continuity from one class meeting to the next was:
  \item The pace of the course was:
  \item The instructor’s assessment of your progress in the course was:
  \item The texts and supplemental learning materials used in the course were:\\
  
Board of regents\\
  \item Description of course objectives and assignments:
  \item Communication of ideas and information:
  \item Expression of expectations for performance:
  \item Availability to assist students in or outside of class:
  \item Respect and concern for students:
  \item Stimulation of interest in the course:
  \item Facilitation of learning:
  \item Overall assessment of instructor:
\end{enumerate}

\label{lastpage}

\end{document}